\documentclass[a4paper]{jpconf}
\usepackage{graphicx}
\usepackage{tikz}
\usepackage{url}

\usepackage[backend=biber,
            doi=true,
            firstinits=true,
            maxnames=10,
            minnames=5,
            sorting=none]{biblatex}
\usetikzlibrary{arrows,shadows, shapes}

\begin{document}
\title{Improved Cloud resource allocation: how INDIGO-DataCloud is overcoming the
current limitations in Cloud schedulers}

\author{
Alvaro Lopez Garcia$^1$,
Lisa Zangrando$^2$,
Massimo Sgaravatto$^2$,
Vincent Llorens$^3$,
Sara Vallero$^4$,
Valentina Zaccolo$^4$,
Stefano Bagnasco$^4$,
Sonia Taneja$^5$,
Stefano Dal Pra$^5$,
Davide Salomoni$^5$,
Giacinto Donvito$^6$
}

\address{
$^1$ IFCA (CSIC--UC), Avda. los Castros s/n, 39005 Santander, Spain\\
$^2$ INFN Padova, Via Marzolo 8, 35131 Padova, Italy \\
$^3$ CC-IN2P3 (CNRS), 21 Avenue Pierre de Coubertin, 69627 Villeurbanne CEDEX, France\\
$^4$ INFN Torino, Via Pietro Giuria 1, 10125 Torino, Italy \\
$^5$ INFN CNAF, Viale Berti Pichat 6/2, 40127 Bologna, Italy \\
$^6$ INFN Bari, Via E. Orabona 4, 70125 Bari, Italy
}

\ead{aloga@ifca.unican.es}

\begin{abstract}
Performing efficient resource provisioning is a fundamental aspect for any
resource provider. Local Resource Management Systems (LRMS) have been used in
data centers for decades in order to obtain the best usage of the resources,
providing their fair usage and partitioning for the users. In contrast, current
cloud schedulers are normally based on the immediate allocation of resources on
a first-come, first-served basis, meaning that a request will fail if there are
no resources (e.g. OpenStack) or it will be trivially queued ordered by entry
time (e.g. OpenNebula). Moreover, these scheduling strategies are based on a
static partitioning of the resources, meaning that existing quotas cannot be
exceeded, even if there are idle resources allocated to other projects. This is
a consequence of the fact that cloud instances are not associated with a
maximum execution time and leads to a situation where the resources are
under-utilized. These facts have been identified by the INDIGO-DataCloud
project as being too simplistic for accommodating scientific workloads in an
efficient way, leading to an under-utilization of the resources, a non
desirable situation in scientific data centers.  In this work, we will present
the work done in the scheduling area during the first year of the INDIGO
project and the foreseen evolutions.
\end{abstract}

\section{Introduction}

The Cloud computing model is being progressively adopted by the scientific
community as a promising paradigm. Cloud infrastructures are now common
in the computing resources portfolio that any scientific resource provider
offers to its user communities. However, in spite of its advantages, the
cloud still has several areas that need to be improved in order to expose to
the researchers its fully functional power. In this context,
INDIGO-DataCloud\footnote{Referred to as INDIGO from now.} (INtegrating
Distributed data Infrastructures for Global ExplOitation)
\cite{salomoni2016indigo}, a project funded under the Horizon 2020 framework
program of the European Union, addresses the challenge of developing an
advanced data and computing platform targeted at scientific communities,
deployable on multiple hardware, and provisioned over hybrid e-Infrastructures.

The project is built by putting together three different layers that
integrate together providing
a scientific computing infrastructure, namely the
infrastructure layer, the platform layer and the user interface layer. In this
paper we will describe how INDIGO tackles the identified gaps in terms of
scheduling and resource provisioning at the Infrastructure as a Service layer,
overcoming the current limitations in the cloud management frameworks adopted
by the project: OpenStack and OpenNebula (ONE).

Although the Cloud model heavily improves the resource usage in the data
centers, there are still open issues in the resource management area. In a
Cloud environment, the concept of ``task'' or ``job'' (that, implicitly has an
associated duration) does not exist. Instead, the Cloud model focuses on
``resources'' (that is, networks, servers, storage, etc.) that are provisioned on
demand by the users. This self-service of resources assumes that there is no
duration associated to them, thus making difficult to perform an efficient
resource provisioning from the provider standpoint.

Moreover, in the existing Cloud Management Frameworks the implemented
scheduling strategy is just based on the immediate First Come First Served
model \cite{Sotomayor2009,Litvinski2013}; consequently a request will be
rejected if there are no resources immediately available: it is up to the
client to later re-issue the request.  It must be also stressed that the
adopted resource allocation model is based on a static partitioning, so that
every user group is assigned an agreed and fixed quota of resources which
cannot be exceeded even if there are idle resources available but allocated to
other groups.

Today the efficient management of Cloud resources is therefore an open issue.
The INDIGO project is addressing it through several, complementary approaches
that will be explained in the upcoming sections. In Section~\ref{sec:sched} we
will explain how we implemented fair-share scheduling strategies in both
OpenStack and OpenNebula. In Section~\ref{sec:preempt} we present an
implementation of preemptible instances for OpenStack clouds. In
Section~\ref{sec:pd} we describe the Partition Director, a tool that esas
the transition of computing nodes between different infrastructures.

\section{Advanced Scheduling}
\label{sec:sched}

On large public cloud providers, in which resources are ideally infinite,
multi-tenant applications are able to scale in and out driven only by their
functional requirements and tenants are normally charged a posteriori for their
resource consumption.

Smaller private Cloud for scientific computing mostly operates in a saturated
regime. In this case, tenants are charged a priori for their computing needs by
paying for a fraction of the computing (or storage) resources. To optimize the
use of the data center an advanced resource allocation policy is needed.

We consider a scenario in which the available resources are divided into
statically and dynamically partitioned. Statically assigned resources are
partitioned among projects according to fixed shares, while dynamical assets
follow the effective requests per project, granting efficient and fair access
to such resources to all projects.

The general topic of advanced resource scheduling is addressed by several
components of the INDIGO project. On the one hand we have implemented fair
share policies by means of the FairShare Scheduler (FaSS) for OpenNebula and
Synergy for OpenStack. On the other hand we have developed an extension for
OpenStack allowing the execution of preemptible instances.

\subsection{Synergy}

Synergy \cite{synergy} has been develeped in the INDIGO-DataCloud project as an
extensible general purpose management service designed for executing tasks in
OpenStack. Its functionalities are provided by a collection of managers which
are specific and independent pluggable tasks executed periodically, like cron
jobs, or interactively through a RESTful API.  Different managers can coexist
and they can interact with each other or with different OpenStack services in a
loosely coupled way.

\begin{figure}[htbp]
    \centering
    \includegraphics[width=0.7\textwidth]{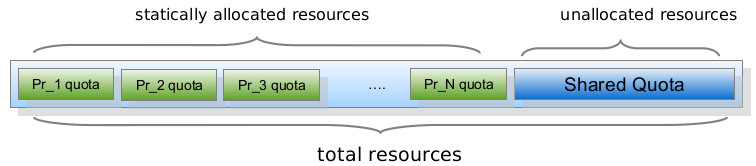}
    \caption{Synergy quota structure.}
    \label{fig:synergy-quotas}
\end{figure}

To enable a more effective and flexible resource allocation and utilization in
OpenStack, six specific managers (as presented in Section~\ref{sec:arch}) have been
implemented. Together they provide an advanced scheduling and resource
allocation strategy based on a fair-share algorithm and a persistent priority
queue. Synergy can maximize the resource utilization by allowing the OpenStack
projects to consume extra shared resources in addition to those statically
assigned. Therefore the projects can access two different kind of quotas: the
private quota and the shared one. On the one hand, the private quota is
composed of resources statically allocated and managed using the ``standard''
OpenStack policies. On the other hand, the shared quota refers to the resources
that can be shared among different projects: it is composed of resources not
statically allocated and they are fairly distributed among the different users
by Synergy. The size of such quota is calculated as the difference between the
total amount of cloud resources and the total resources allocated to the
private quotas (Figure~\ref{fig:synergy-quotas}).

Only the projects selected by the administrator can consume the shared
resources and, unlike the private quota, the user requests that cannot be
immediately satisfied are not rejected but instead inserted in a persistent
priority queue. The priority is periodically recalculated by Synergy by
enforcing specific fair-share policies defined by the administrator which
consider the historical resource usage in a well defined time window,
the project and user shares and some specific weights (e.g. age, decay, cpu
usage, memory).

The chosen fair-share algorithm for Synergy is the Priority Multifactor
implemented by the SLURM batch scheduler \cite{slurm-multifactor}. The enqueued
requests are processed according to their priority ordering. In case there are
no resources available for the selected request, it is skipped and Synergy will
process the next one in the queue (i.e. backfilling): this allows to maximize
the resource utilization and, at the same time, avoid to block the queue for
long time.

\subsubsection{Architecture}
\label{sec:arch}

\begin{figure}[htbp]
    \centering
    \includegraphics[width=0.5\textwidth]{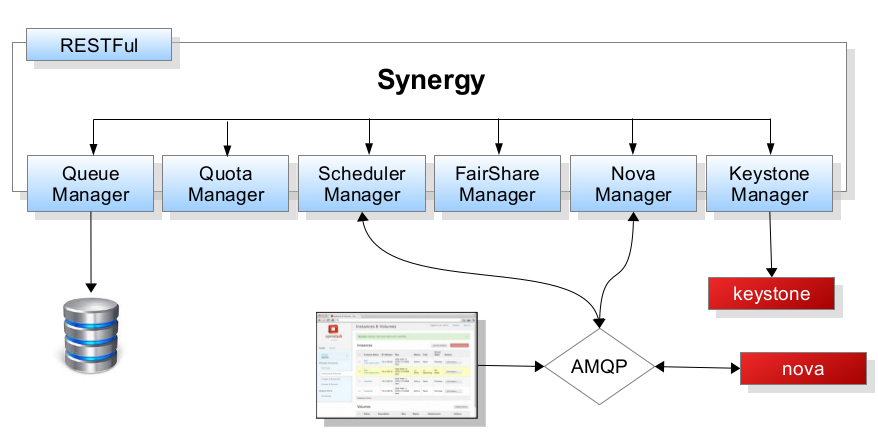}
    \caption{Synergy's architecture.}
    \label{fig:synergy-arch}
\end{figure}

Synergy's architecture is depicted in Figure~\ref{fig:synergy-arch}. The
Nova-Manager intercepts all incoming user requests. The FairShare-Manager
calculates and assigns to each of them the proper priority value. Such requests
are immediately inserted in a persistent priority queue by the Queue-Manager.
The role of the Scheduler-Manager is to process the user requests by fetching
them from the queue according to their priority based ordering. The selected
request is finally sent to the OpenStack Compute Scheduler through
Synergy's Nova-Manager. In the scenario of full resource utilization for a
specific quota, the Quota-Manager advises the Scheduler-Manager to wait until
the compute resources become available again. In case of failure, the
Scheduler-Manager provides a retry mechanism which handles the failed
requests by inserting them again into the queue (up to a given number of
retries). The priority of the queued requests is periodically
recalculated in order to increase the priority of older requests.

To prevent any possible interaction issue with the existing OpenStack clients,
no new states have been added so that from the client point of view the queued
requests remain in ``Scheduling'' state till the compute resources are
available.

Whenever applicable Synergy has been developed by adopting the full OpenStack
style according to the specific development guidelines, including the coding
style and the technologies to be used: the source code is in launchpad
\cite{lp:synergy-service,lp:synergy-manager} and the OpenStack Continuous
Integration system is used. Synergy is a full python service accessed by
clients through a well defined RESTful API which provides either management
commands for handling the plugged managers (e.g.  activate manager, suspend,
get status, etc) or executing specific commands on the selected manager (e.g.
get/set quota, list queues, etc). The API can be extended for including new
specific interfaces. For what concerns the internal interfaces, Synergy
interacts with the OpenStack services as Nova and Keystone, by invoking the
exposed RESTful interfaces or using the internal RPC APIs. All Synergy default
configuration options are defined in the \texttt{synergy.conf} file.

\subsection{FaSS: A fair-share scheduler for OpenNebula}

The current OpenNebula scheduler is first-in-first-out (FIFO) and it is
based only on static resources partitioning among the projects, which is not
efficient for a scientific datacenter. FaSS grants fair access to dynamic
resources prioritizing tasks assigned according to an initial weight and to the
historical resource usage, irrespective of the number of tasks running
on the system.

The software was designed to be fully synchronized with the ONE authentication,
quota, monitoring and accounting systems, without being intrusive in the ONE
code and keeping to the minimum the dependencies on the ONE implementation
details. In such a way, the code is fairly independent on future ONE releases.

The ONE original scheduler is kept to match the resources to requests and FaSS
is structured as a self-contained module interacting only with the ONE XML-RPC
interface.

\subsubsection{Architecture}

\begin{figure}[htbp]
    \centering
    \includegraphics[width=0.5\textwidth]{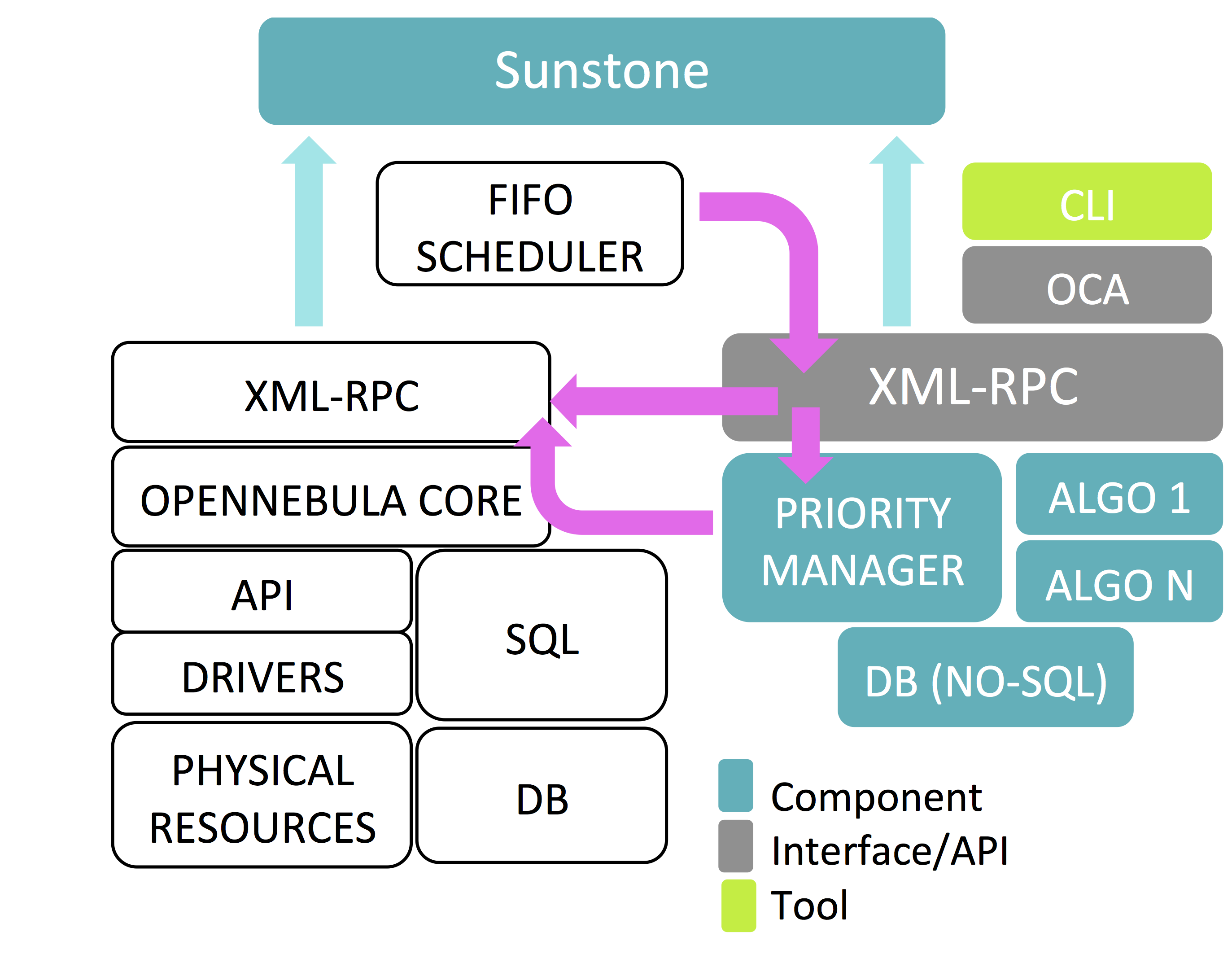}
    \caption{FaSS architecture. }
    \label{fig:fass}
\end{figure}

The architecture of the FaSS service is depicted in Figure~\ref{fig:fass}. It
has been conceived following the ONE conceptual design \cite{web:onearch}. In
the figure, white blocks are native ONE components, while the newly developed
components are shown in blue. New tools are pictured in green, and
interfaces/API in gray. Purple arrrows represent internal component
interactions. Blue arrows represent the API-Sunstone GUI interaction.

The main component of FaSS is the Priority Manager. Its task is to
calculate periodically priorities for queued jobs. It receives input data from
the ONE core through XML-RPC calls, and from the Priority Database
through a no-SQL interface. This choice has been made in order to avoid storing
large XML strings, which could easily be corrupted in a single SQL entry.

The Priority Database holds the modules internal data: initial priority
values, historical information on resource usage and recalculated priority
values. The Priority Manager interacts with a set of pluggable algorithms, such as
MultiFactor \cite{slurm-multifactor} and FairTree \cite{slurm-fair-tree}, to
calculate priorities. It exposes an XML-RPC interface, independent from the one
exposed by the ONE core, and uses the Priority Database as data back-end.

The second fundamental component of FaSS is the XML-RPC interface of the
Priority Manager, which replaces the ONE XML-RPC interface in the communication
with the FIFO scheduler.  The queue of pending jobs to be processed must be the
one ordered according to the priorities calculated by the Priority Manager. The
XML-RPC server, then, catches the scheduler calls, provides a reordered queue
and redirects unknown methods to the core of ONE XML-RPC server.

The Sunstone web-based GUI service is extended to allow monitoring and
operating the FaSS module functionalities. The Priority Manager will be
accessed by clients, interfacing to the module core through a set of bindings
analogous to the ONE Cloud API (OCA) \cite{web:oneapis}.

A prototype of the scheduler with limited functionalities is ready
\cite{web:fass}. Namely, the prototype comprises the Priority Manager and
its XML-RPC interface. For the next INDIGO project release, a fully featured
prototype, including some algorithms and installation tools, will be made
available.

\subsection{Preemptible Instances support}
\label{sec:preempt}

Resources in a cloud environment do not
have an associated lifespan prevents resource providers to perform advanced
scheduling strategies (like backfilling) to increase their infrastructure
utilization. Implementing fair share policies (as depicted in
Section~\ref{sec:sched}) can alleviate the infrastructure utilization problem.
Another, complementary, approach is the implementation of preemptible
instances, that is instances that can be terminated  whenever the resource
provider decides so.

In commercial cloud world, providers are offering this functionality to their
users, but there is not a similar implementation in the open source Cloud
Management Frameworks. For
example, in the Amazon EC2 Spot Instances \cite{amazon2015spot} users are able
to select how much they are willing to pay for their resources in a market
where the price fluctuates accordingly to the demand. Whenever the reference
price goes above the user bid, those requests are terminated.
The Google Cloud Engine (GCE) \cite{web:googlecloud} has released a new product
branded as \emph{Preemptible Virtual Machines} \cite{web:googlecloud:preempt},
where machines are limited to 24h of execution and can be terminated without
prior notice.

Resource providers can offer these resources to the users at a fraction of the
original price, increasing their resource utilization without preventing higher
priority instances to be executed (since the preemptible instances would be
terminated whenever needed). Users would get access to a
higher number of resources that they can use for fault-tolerant tasks like
batch processing.
We have proposed this concept to be included in the upstream OpenStack Compute
component \cite{opie-proposal}, and an implementation (described below) is
available \cite{opie}.

\subsubsection{OPIE: OpenStack Preemptible Instances Extension}

We have developed OPIE \cite{opie} as an extension for OpenStack, allowing
to execute preemptible instances in an existing infrastructure. In the current
implementation, preemptible instances can run within the user quota and as far
as there are not preventing normal requests from being executed. Whenever the
scheduler detects that a normal instance cannot be executed because of a
preemptible instance, it triggers its termination, according to several filter
and weight functions, configurable by the resource provider.
OPIE is composed of the following modules:

\begin{itemize}
    \item A modified scheduler, that augments the OpenStack FilterScheduler
        with the ability to execute normal and preemptible instances,
        configurable and extensible via additional filter and raking functions
        so that each resource provider can implement its policies.

	\item A special host manager, able to report different usage values so that
        the OPIE scheduler is able to take the appropriate decision.

    \item A plugin for the OpenStack clients that allows booting preemptible
        instances in a cloud infrastructure with OPIE enabled.
\end{itemize}

\section{Partition Director}
\label{sec:pd}

A typical Grid scientific computing centre is organized and dimensioned towards
a quite specific usage pattern.  Such computing resources are normally managed
by a batch system according to some sort of fair share policy so that a convenient resource usage
is enforced. User communities are now asking for the provisioning of resources
exposing a Cloud interface from such infrastructures. Examples of some relevant
use cases are:

\begin{itemize}
    \item A group would like to dedicate a subquota of its
        pledges at the centre to a ``cloud computation campaign'', then
        eventually go on as usual with batch mode.

    \item A group wants to gradually migrate its computing resources to the cloud, few
        at a time.

    \item One user group needs some resources for interactive usage, which
        currently is handled by providing powerful hosts ``User Interface'',
        bypassing the official pledge count.
\end{itemize}

The Partition Director (PD) is a tool developed to offer the ability to convert
a set of Worker Nodes in the batch cluster (currrently LSF) to Cloud compute
nodes (currently OpenStack), and vice versa.  The tool can be used by a site
administrator, but can also be driven by a user group: it can trigger node
conversion to add more resources toward a partition with many more requests.
The remainder of this section describes its functionalities and working
principles.

\subsection{Functionality}

The Partition Director eases the management of a hybrid data center providing
both batch system and cloud based services.  Partition Director provides site
administrators with an instrument to dynamically resize sub-quotas of a given
user group among different infrastructures (HPC, Grid, local batch systems,
cloud resources).  Physical computing resources, in fact, can work as member of
batch system cluster or as compute node on cloud resources.

At its basic level, the PD is in charge of switching the role of selected
computing machines (Worker Nodes) from a batch cluster to a cloud one, enabling
them as Compute Node (CN) and vice versa, managing the intermediate transition
states, thus ensuring consistency, Moreover, it adjusts the shares at the batch
side, to enforce an overall uniform quota to groups using both cloud and batch
resources.

\subsubsection{Node transitions}

\begin{figure}
    \begin{center}
        \begin{tikzpicture}[
          scale=0.5, every node/.style={transform shape},
          node distance=15ex,>=stealth',shorten >=.95ex,%
          every loop/.style={->,shorten >=.5ex}]
          \tikzstyle{every node}=[%
           fill=green!50!black!20,%
           draw=green!50!black,%
           minimum size=10ex,%
           circle,%
           thick%
          ]

          \node[fill=pink,radius=0.75] (B) {$B:1$};
          \node[fill=pink] (B2CR) [right of=B] {$B2CR:1$};
          \node (B2C) [right of=B2CR] {$B2C:2$};
          \node (C2B) [below of=B] {$C2B:2$};
          \node (C2BR) [right of=C2B] {$C2BR:2$};
          \node (C) [below of=B2C] {$C:2$};

          \path [thick]
           (B)  edge [loop left] (B)

           (B2C) edge [loop right] (B2C)
           (C2B) edge [loop left] (B2C)
           (C)  edge [loop right] (C)
           (B)  [<->] edge (B2CR)
           (B2CR) [->] edge (B2C)
           (C2BR)  [->] edge (C2B)
           (C) [<->] edge (C2BR)
           (B2C) [->] edge (C)
           (C2B) [->] edge (B)
          ;
          ;
        \end{tikzpicture}
    \end{center}
    \caption{The Status Transition Map. Each node publishes a dynp ``load
        index'' with value 1 (pink, available for batch jobs) or 2 (green, no
        new batch tasks). The batch system alters each new job to require nodes
        having {\tt dynp==1}.  When a transition from batch ($B$) or cloud
        ($C$) is triggered for a node, it goes through different states: $B2CR$
        or $C2BR$ for validation, $B2C$ or $C2B$ for Draining tasks of the
        partition being left, $C$ or $B$ to become active in the new partition.}
    \label{fig:statemachine}
\end{figure}
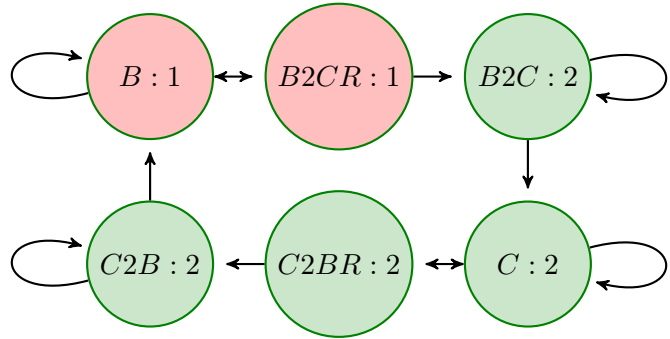

The role conversion of nodes between Batch and Cloud (WN $\leftrightarrow$ CN),
are managed by a finite state machine as represented in
Figure~\ref{fig:statemachine} exploiting a partitioning concept \cite{MCORE}
implemented as a {\em partition driver} script. Each physical machine in the
cluster has one and only one status at any point in time and only two of them
are stable: $B$ for nodes active as WN and $C$ for those active as CN.
Transitory states are intended for validation or draining phases: the
validation step is required to evaluate the consistency of each transition
request; the draining phase is needed to ensure that a node is actually free
and clean before activating it with the new role. It is worth noting that
draining from cloud to batch requires us to define a time limit, because there
is no maximum lifetime for VMs running in a CN as there is for jobs running in
a WN. To address this problem, the Partition Director sets a Time To live (TTL)
value as defined by the Machine/Job Features Task Force
\cite{alef2016machine}. VMs can check whether a TTL is defined and stop
themselves gracefully. After the TTL has expired, remaining VMs are destroyed
and the role conversion can complete.

\subsubsection{Share adjustment}

Whenever nodes are moved from the partition they belong to, the size of both
the batch and cloud cluster changes. However, when resources are moved from
batch to cloud, they are assigned to a single tenant (user group). The
consequence is that the shares are to be rebalanced at the batch system side,
to guarantee that at the overall pledges of all the user communities working
with the centre are unaffected.

\section{Conclusions}

INDIGO has tackled the existing cloud resource provisioning and scheduling
limitations in different, complementary ways. We have focused
on the enhancement of the scheduling policies through the development of
services that makes it possible to deliver fair-share policies and
through the introduction of preemptible instances. Furthermore we have
developed a tool that eases the transition of nodes from one infrastructure
managed by a LRMS to a Cloud one, preserving the user share and quota.

These components have been deployed at several production sites (like
INFN-Padova, INFN-CNAF in Italy and IFCA-CSIC in Spain)  in order to validate
its functionalities, its robustness as well as its stability under different
real usage conditions. Some test results have confirmed the expected limitation
of the SLURM MultiFactor algorithm as documented at \cite{slurm-multifactor},
thus Synergy developers are working on the integration of the SLURM FairTree
\cite{slurm-fair-tree}, a more sophisticated algorithm which fully solves the
observed issue.

\section{Acknowledgments}

The authors want to acknowledge the support of the INDIGO-DataCloud (grant
number 653549) project, funded by the European Commission's Horizon 2020
Framework Programme.

\printbibliography
%\bibliography{references}   % name your BibTeX data base

\end{document}